%% file: main.tex
\documentclass{article}
\usepackage{graphicx} 
\usepackage{amsmath} 
\usepackage{amsthm}
\usepackage{amsfonts}
\usepackage{authblk}
\usepackage{subcaption}
\usepackage{subfiles}
\usepackage{xcolor}
\usepackage{booktabs}
\usepackage{geometry}
\usepackage{placeins} 
\usepackage{makecell}
\usepackage{hyperref}
\usepackage{multirow}
\usepackage{mathtools}
\usepackage{amssymb}
\usepackage{bm}
\usepackage{float} 
\usepackage[authoryear]{natbib}

\usepackage{microtype}      
\title{FLEX-CP-DT: A Flexible Conditional Power Framework for Interim Futility Analysis in Clinical Trials with Count Endpoints and Temporal Trends}
\author[1]{Yanzhao Wang \thanks{Corresponding author: \texttt{yzwang1994@gmail.com}}}
\author[1]{Dateng Li}
\author[2]{Ningya Wang}
\author[1]{Haitao Gao}
\author[1]{Chenguang Wang}
\affil[1]{Biostatistics and Data Management, Regeneron Pharmaceuticals, Tarrytown, NY}
\affil[2]{Department of Biostatistics and Epidemiology, Rutgers University, New Brunswick, NJ}
\date{}

\begin{document}
 
\maketitle
\begin{abstract}
 Many large-scale phase III trials with recurrent event endpoints include a pre-planned interim analysis to evaluate early futility. Conditional power (CP), which quantifies the probability of achieving statistical significance at the final analysis given the interim data, is a commonly used tool to support such decisions. The standard negative binomial model with an offset term, widely adopted for analyzing recurrent events, implicitly assumes that event rates and treatment effects remain constant over the study period. At the interim analysis, however, a substantial proportion of patients have incomplete follow-up, and when the treatment effect is delayed in onset or diminishes over time, the constant-rate assumption introduces systematic bias into the interim estimate and can lead to incorrect futility decisions. In this paper, we propose FLEX-CP-DT, a piecewise negative binomial framework that captures temporal trends in both event rates and treatment effects without imposing the constant-rate assumption. The framework yields a formula-based conditional power calculation that does not require resampling or trial simulation at the interim stage. Through extensive simulations spanning constant-effect and delayed-onset scenarios, we demonstrate that FLEX-CP-DT performs comparably to the standard approach when the constant-rate assumption holds and improves interim futility decision-making when it is violated. A case study calibrated to a published phase 3 bronchiectasis trial further illustrates the practical advantage of the proposed method in reducing the probability of falsely terminating an efficacious drug with delayed treatment onset.
\end{abstract}

\subfile{Sections/introduction}

\subfile{Sections/methodology}

\subfile{Sections/simulation}

\subfile{Sections/application}
\subfile{Sections/discussion}


\section*{Author Contributions}
\label{AuthorContributions}

Yanzhao Wang and Dateng Li contributed equally to this work.
\textbf{Yanzhao Wang}: contributed to methodology, simulation, data analysis, and investigation.
\textbf{Dateng Li}: contributed to conceptual ideas, methodology, investigation, and supervision.
\textbf{Ningya Wang}: contributed to simulation and data analysis.
\textbf{Haitao Gao}: contributed to conceptual ideas, methodology, and supervision.
\textbf{Chenguang Wang}: contributed to conceptual ideas, methodology, and supervision.
All authors contributed to the writing and editing of the original draft.

\section*{Data Availability Statement}
\label{DataAvailability}

The data that support the findings of this study are available from the corresponding author upon reasonable request.

\newpage
\bibliographystyle{agsm}
\bibliography{Sections/Reference}
 
\end{document}

%% file: Sections/introduction.tex
\section{Introduction}
\label{Introduction}

Many long-term and large-scale phase III trials have a pre-planned formal unblinded interim analysis \citep{FDA2019, Walter2020}. Results from the interim analysis can be used for early decision-making, including potential trial stopping for futility to save resources \citep{Proschan1995, brannath2002recursive}. 

Conditional power (CP) and predictive power of success (PPoS) are commonly used statistical approaches to support decision-making at interim analyses. Conditional power, first proposed by Halperin et al.\ \cite{Halperin1982}, quantifies the probability of study success given the interim results while assuming a fixed effect size for the remainder of the trial. \cite{Lan1988} generalized the CP calculation using the B-value decomposition, which exploits the independent increment structure of test statistics. PPoS, introduced by \cite{Spiegelhalter1986}, averages the conditional probability of success over the predictive distribution of the treatment effect, incorporating uncertainty in the effect size rather than conditioning on a single fixed value. Both CP and PPoS have been developed for continuous, binary, and time-to-event endpoints \citep{kundu2021conditional}.

Recurrent event or count data endpoints are frequently used as primary endpoints in certain disease areas. Examples include exacerbation counts in chronic obstructive pulmonary disease (COPD) and asthma trials, heart failure hospitalizations in cardiovascular disease, and relapses in multiple sclerosis studies \citep{keene2008statistical, Nicholas2011, Rogers2014, Friede2010}. These endpoints are commonly analyzed using a negative binomial (NB) regression model with an overdispersion parameter.  Unlike the Poisson model, which assumes the variance equals the mean, the NB model
accommodates a variance larger than the mean (overdispersion) \citep{Cameron2013}. There have been developments in trial design for count endpoints based on the NB model, including fixed designs \citep{Zhu2013}, blinded sample size re-estimation \citep{Asendorf2019, Zapf2020}, and group sequential designs \citep{Mutze2018}. More recently, Quan et al.\ \cite{Quan2024} extended this work to two-stage adaptive designs, providing conditional power calculation and sample size re-estimation for count endpoints based on unblinded interim results.



Most existing approaches for interim futility analysis with count endpoints employ a negative binomial (NB) model with an offset term to adjust for varying follow-up durations across patients. This implicitly assumes that both event rates and treatment effects remain constant throughout the study period.  These concerns are amplified at an interim analysis, where a delayed onset or attenuation of the treatment effect can yield a treatment effect estimate that is unrepresentative of the cumulative effect at the planned end of study, potentially leading to incorrect futility decisions. \cite{Quan2021} acknowledged the potential impact of delayed treatment effects on interim decision-making and proposed exploratory analyses to assess temporal patterns, but did not incorporate the temporal trend into the conditional power calculation. To the best of our knowledge, no existing method directly captures temporal trends in event rates and treatment effects for interim futility analysis with count-data endpoints.

In this study, we propose FLEX-CP-DT, a novel approach that captures temporal trends in both event rates and treatment effects over the study period. Unlike the standard NB model with an offset term, FLEX-CP-DT does not impose the assumption of constant rates, enabling more accurate estimation of the treatment effect at the interim analysis. The rest of the paper is organized as follows. In Section~\ref{Methodology} we describe the proposed FLEX-CP-DT approach. In Section~\ref{Simulation} we present simulation studies to assess performance. In Section~\ref{Application} we apply the proposed method to a real clinical trial. In Section~\ref{Discussion} we provide discussion and concluding remarks.

%% file: Sections/methodology.tex
\section{Methodology}
\label{Methodology}

\subsection{Standard negative binomial model}
Consider a randomized clinical trial comparing an investigational treatment ($r=1$) against a control ($r=0$) with respect to a recurrent event endpoint. The key notation is summarized in Table~\ref{Methodology_model data introduction}.

\begin{table}[!htb]
\centering
\caption{Notation for a clinical trial with a recurrent event endpoint.}
\begin{tabular}{lll}
\toprule
Notation & Description & Range \\
\midrule
$n_r$ & Number of subjects in arm $r$ & $r = 0,1$\\
\midrule
$T$ & Planned maximum treatment duration  & \\
\midrule
$t_{rs}$  & Observed treatment duration for subject $s$ in arm $r$ &  $s = 1,\ldots,n_r$; $t_{rs}\in [0,T]$\\
\midrule
$y_{rs}$ & \makecell[l]{Observed event count for subject $s$ in arm $r$\\ over the interval $[0,t_{rs}]$}& \\
\bottomrule
\end{tabular}
\label{Methodology_model data introduction}
\end{table}

The negative binomial regression model \citep{Zhu2013} specifies
\begin{equation}
\begin{split}
    & Y_{rs}\mid\mu_{rs},a\sim \text{NegBin}(\mu_{rs},a),\\
    &\Pr(Y_{rs} = y_{rs}\mid\mu_{rs},a,t_{rs}) = \frac{\Gamma(a^{-1} + y_{rs})}{\Gamma(a^{-1})\, y_{rs}!}\left(\frac{a\mu_{rs}}{1+a\mu_{rs}}\right)^{y_{rs}}\left(\frac{1}{1+a\mu_{rs}}\right)^{a^{-1}},\\
    &\log(\mu_{rs}) = \log(t_{rs}) + \beta_0 + \beta_1\cdot r,
\end{split}
\label{Methodology_NB model}
\end{equation}
where $\mu_{rs}$ is the expected number of events for subject $s$ in arm $r$ over duration $[0,t_{rs}]$, $a>0$ is the dispersion parameter, and the offset $\log(t_{rs})$ adjusts for variable exposure duration. The variance takes the form $\text{Var}(Y_{rs}\mid\mu_{rs},a) = \mu_{rs}+a\mu_{rs}^2$. Here $\beta_0$ represents the log annualized event rate in the control arm and $\beta_1$ represents the log event rate ratio between treatment and control arms, so that $e^{\beta_1}$ is the rate ratio of primary interest.

The joint log-likelihood function is
\begin{equation}
\begin{split}
    &L(\beta_0,\beta_1,a\mid\{y_{0s}\}_{s=1}^{n_0},\{y_{1s}\}_{s=1}^{n_1},\{t_{0s}\}_{s=1}^{n_0},\{t_{1s}\}_{s=1}^{n_1}) \\
    &= \sum_{r=0}^1\sum_{s=1}^{n_r}\left[\log\frac{\Gamma(a^{-1} + y_{rs})}{\Gamma(a^{-1})\,y_{rs}!}+y_{rs}\log(a\mu_{rs})-(y_{rs}+a^{-1})\log(1+a\mu_{rs})\right],
\end{split}
\label{Methodology_NB model joint likelihood}
\end{equation}
and maximum likelihood estimation (MLE) yields the point estimate $\hat{\beta}_1$ along with its standard error for hypothesis testing or constructing confidence interval.

\subsubsection{Constant-rate assumption and its implication}
A fundamental assumption embedded in model~\eqref{Methodology_NB model} is that both $\beta_0$ and $\beta_1$ are constant, implying that the event rate within each arm and their ratio $e^{\beta_1}$ remain constant over the entire treatment period. In practice, however, this assumption may be violated when the treatment effect exhibits temporal patterns; for example, a delayed onset effect or a diminishing effect over time. Under such scenarios, the expected number of events for subject $s$ in arm $r$ over $[0,t]$ should be expressed as
\begin{equation}
\mu_{rs}(t)= \int_0^t\exp\left\{\beta_0(t^\prime)+\beta_1(t^\prime)\cdot r\right\}dt^\prime,
\label{Methodolog_cumulative events}
\end{equation}
where $\beta_0(t^\prime)$ and $\beta_1(t^\prime)$ denote the time-varying log event rate and log rate ratio at time $t^\prime$, respectively.

Despite such violations, the standard negative binomial model remains appropriate for the final analysis when nearly all subjects complete the planned treatment duration $T$. To see this, define the arm-specific cumulative mean rate over $[0,T]$ as
  \begin{equation}
  \Lambda_r = \int_0^T\exp\{\beta_0(t^\prime)+\beta_1(t^\prime)\cdot r\}\,dt^\prime = \mu_r(T),\quad r=0,1.
  \label{Methodolog_cumulative_mean_rate}
  \end{equation}
  When $t_{rs} \approx T$ for all subjects, we have $\mu_{rs}(t_{rs}) \approx \Lambda_r\cdot \frac{t_{rs}}{T}$, and the log-mean can be written as
  \begin{equation}
  \log\{\mu_{rs}(t_{rs})\} = \log(\Lambda_r) + \log(t_{rs}) -\log(T)= \log(\Lambda_0/T) + \log(t_{rs}) + \log(\Lambda_1/\Lambda_0)\cdot r.
  \label{Methodolog_NB overall efficacy estimation}
  \end{equation}
  Comparing~\eqref{Methodolog_NB overall efficacy estimation} with model~\eqref{Methodology_NB model}, we identify $\beta_0 = \log(\Lambda_0/T)$ and $\beta_1 = \log(\Lambda_1/\Lambda_0)$, so that $e^{\beta_1} = \Lambda_1/\Lambda_0 = \mu_1(T)/\mu_0(T)$. That is, the treatment effect estimated by the standard model is interpretable as the ratio of cumulative mean event counts over the planned treatment period.

Throughout this paper, we denote $\beta^*_0 = \log(\Lambda_0/T)$ and $\beta^*_1 = \log(\Lambda_1/\Lambda_0)$ as the target estimands for the final analysis.

\subsubsection{Limitation at the interim analysis}
Timing of a protocol specified interim analysis typically occurs after a specified number of subjects have completed a minimum follow-up duration. Let $\mathcal{D}_\text{IA}$ and $\mathcal{D}_\text{FA}$ denote the analysis sets at the interim and final analyses, respectively. Note that, subjects in $\mathcal{D}_\text{IA}$ have heterogeneous actual treatment durations $t_{rs}$, with a substantial proportion contributing only partial follow-up.

Consider the method-of-moments (MoM) estimator \citep{wang2019sample} of the rate ratio at each analysis:
\begin{equation*}
\begin{split}
    &\exp(\hat{\beta}_{1I}^{\text{MoM}}) = \frac{\sum_{\{r=1,s\}\in\mathcal{D}_\text{IA}}y_{1s}/\sum_{\{r=1,s\}\in\mathcal{D}_\text{IA}}t_{1s}}{\sum_{\{r=0,s\}\in\mathcal{D}_\text{IA}}y_{0s}/\sum_{\{r=0,s\}\in\mathcal{D}_\text{IA}}t_{0s}},\\
    &\exp(\hat{\beta}_{1F}^{\text{MoM}}) = \frac{\sum_{\{r=1,s\}\in\mathcal{D}_\text{FA}}y_{1s}/\sum_{\{r=1,s\}\in\mathcal{D}_\text{FA}}t_{1s}}{\sum_{\{r=0,s\}\in\mathcal{D}_\text{FA}}y_{0s}/\sum_{\{r=0,s\}\in\mathcal{D}_\text{FA}}t_{0s}}.
\end{split}
\end{equation*}
Let $\lambda_r(t) = \exp\{\beta_0(t)+\beta_1(t)\cdot r\}$ denote the instantaneous event rate in arm $r$ at time $t$, and let $N_r(t) = |\{s: t_{rs}\ge t\}|$ be the number of subjects in arm $r$ with follow-up at least $t$. Then the MoM estimator in arm $r$ targets, in expectation,
\begin{equation}
\frac{\sum_{s=1}^{n_r}E[y_{rs}]}{\sum_{s=1}^{n_r}t_{rs}} = \frac{\sum_{s=1}^{n_r}\int_0^{t_{rs}}\lambda_r(t^\prime)\,dt^\prime}{\sum_{s=1}^{n_r}t_{rs}}  = \frac{\int_0^{T}\lambda_r(t^\prime)\,N_r(t^\prime)\,dt^\prime}{\int_0^{T}N_r(t^\prime)\,dt^\prime} = \int_0^{T}\lambda_r(t^\prime)\,w_r(t^\prime)\,dt^\prime,
\label{methodology_MoM_weighted_avg}
\end{equation}
where $w_r(t^\prime) = N_r(t^\prime)\big/\int_0^T N_r(u)\,du$ is a weight function satisfying $\int_0^T w_r(t^\prime)\,dt^\prime=1$. Under the constant-rate assumption, $\lambda_r(t^\prime)\equiv e^{\beta_0+\beta_1\cdot r}$ is independent of $t^\prime$, and \eqref{methodology_MoM_weighted_avg} reduces to $e^{\beta_0+\beta^*_1\cdot r}$ regardless of the weight function. Consequently, $\exp(\hat{\beta}_{1I}^{\text{MoM}})$ and $\exp(\hat{\beta}_{1F}^{\text{MoM}})$ target the same quantity $e^{\beta_1}$.

When $\lambda_r(t^\prime)$ varies over time, the estimand in \eqref{methodology_MoM_weighted_avg} depends on $w(t^\prime)$. At the final analysis, $t_{rs}\approx T$ for all subjects, so $N_r(T)\approx n_r$ and $w(T)\approx 1/T$ (uniform weighting), yielding the cumulative rate $\mu_r(T)/T$ and rate ratio $e^{\beta^*_1}$ as desired. At the interim, $N_r(t^\prime)$ is monotonically decreasing in $t^\prime$ due to staggered enrollment, placing disproportionate weight on the early period. The MoM rate ratio at the interim therefore targets

\begin{equation*}
\frac{\int_0^T\lambda_1(t^\prime)\,w_I(t^\prime)\,dt^\prime}{\int_0^T\lambda_0(t^\prime)\,w_I(t^\prime)\,dt^\prime} \;\neq\; e^{\beta^*_1},
\end{equation*}
where $w_I(t^\prime)$ is the interim-specific weight function. 

To address this, we propose the FLEX-CP-DT framework, which accommodates temporal trends in a formula-based CP calculation.

\subsection{The FLEX-CP-DT framework}
 \subsubsection{Piecewise-constant rate model}
To capture temporal trends in event rates, we partition the treatment period $[0,T]$ into $K\ge 2$ intervals at pre-specified change points $0=T_0<T_1<\cdots<T_K=T$, and specify piecewise-constant functions:
  \begin{equation}
      \beta_0(t)=
  \begin{cases}
  \beta_{0}^{(1)}, & t \in (0,T_1]\\
  \beta_{0}^{(2)}, & t \in (T_1,T_2]\\
   \vdots\\
  \beta_{0}^{(K)}, & t \in (T_{K-1},T_K]
  \end{cases},\quad\beta_1(t)=
  \begin{cases}
  \beta_{1}^{(1)}, & t \in (0,T_1]\\
  \beta_{1}^{(2)}, & t \in (T_1,T_2]\\
   \vdots\\
  \beta_{1}^{(K)}, & t \in (T_{K-1},T_K]
  \end{cases}
  \label{Methodology_piecewise event rates}
  \end{equation}
where $\beta_0^{(k)}$ and $\beta_1^{(k)}$ denote the log annualized event rate and log rate ratio on the $k$th interval $(T_{k-1},T_k]$, respectively. Define $K^\prime(r,s)\in\{1,\ldots,K\}$ as the index of the interval containing the end of subject $(r,s)$'s follow-up, i.e., $t_{rs}\in(T_{K^\prime(r,s)-1},\,T_{K^\prime(r,s)}]$. 

Let $y_{rs}^{(k)}$ denote the number of events observed from subject $s$ in arm $r$ during the $k$th interval, so that $y_{rs}=\sum_{k=1}^{K^\prime(r,s)}y_{rs}^{(k)}$. The interval-specific expected count is
  \begin{equation}
  \mu_{rs}^{(k)}=\begin{cases}
      (T_{k}-T_{k-1})\cdot e^{\beta_{0}^{(k)}+r\cdot\beta_{1}^{(k)}},& k =1,\ldots,K^\prime(r,s)-1,\\
      (t_{rs}-T_{K^\prime(r,s)-1})\cdot e^{\beta_{0}^{(K^\prime(r,s))}+r\cdot\beta_{1}^{(K^\prime(r,s))}},&k=K^\prime(r,s).
  \end{cases}
  \label{Methodology_interval_means}
  \end{equation}

To account for over-dispersion, we introduce subject-specific gamma frailties $\nu_{rs}\stackrel{\text{i.i.d.}}{\sim}\text{Gamma}(a^{-1},a^{-1})$ with $E(\nu_{rs})=1$ and $\text{Var}(\nu_{rs})=a$. Conditional on $\nu_{rs}$, the event process within each interval is a homogeneous Poisson process with rate $\nu_{rs}\cdot e^{\beta_0^{(k)}+r\cdot\beta_1^{(k)}}$, so that the interval-specific counts are conditionally independent:
  \begin{equation}
  Y_{rs}^{(k)}\mid\nu_{rs} \;\sim\; \text{Poisson}\!\left(\mu_{rs}^{(k)}\,\nu_{rs}\right),\quad k=1,\ldots,K^\prime(r,s).
  \label{Methodology_conditional_Poisson}
  \end{equation}

\subsubsection{Marginal likelihood}
Since the interval-specific counts are conditionally independent Poisson given $\nu_{rs}$, the conditional joint PMF for subject $(r,s)$ is
  \begin{equation*}
  f\!\left(y_{rs}^{(1)},\ldots,y_{rs}^{(K^\prime)}\mid\nu_{rs}\right) = \prod_{k=1}^{K^\prime(r,s)}\frac{(\mu_{rs}^{(k)}\,\nu_{rs})^{y_{rs}^{(k)}}\,e^{-\mu_{rs}^{(k)}\,\nu_{rs}}}{y_{rs}^{(k)}!}.
  \end{equation*}
Integrating out $\nu_{rs}\sim\text{Gamma}(a^{-1},a^{-1})$, we obtain the marginal likelihood contribution of subject $(r,s)$ in closed form:
    \begin{equation}
  l(r,s) = \underbrace{\frac{\Gamma(y_{rs}+a^{-1})}{\Gamma(a^{-1})\,y_{rs}!}\left(\frac{a\,M_{rs}}{1+a\,M_{rs}}\right)^{y_{rs}}\left(\frac{1}{1+a\,M_{rs}}\right)^{a^{-1}}}_{\text{NB}(y_{rs};\,M_{rs},\,a)}\;\times\;\underbrace{\frac{y_{rs}!}{\prod_{k}y_{rs}^{(k)}!}\prod_{k=1}^{K^\prime}\left(\frac{\mu_{rs}^{(k)}}{M_{rs}}\right)^{y_{rs}^{(k)}}}_{\text{Multinomial}(y_{rs}^{(1)},\ldots,y_{rs}^{(K^\prime)}\mid y_{rs};\,p_k=\mu_{rs}^{(k)}/M_{rs})},
  \label{Methodology_likelihood_factorization}
  \end{equation}
where $M_{rs}=\sum_{k=1}^{K^\prime(r,s)}\mu_{rs}^{(k)}$ is the total expected count over $[0,t_{rs}]$. It is noteworthy that $l(r,s)$ can be factorized into two components.  The first component corresponds to the likleihood function of a standard negative binomial distribution with mean $M_{rs}$ and dispersion $a$. The second component can be viewed as  the likelihood function of a  multinomial distribution, which leverages the temporal allocation of events across time intervals, which is essential for making $\beta_0^{(k)}$ and $\beta_1^{(k)}$ identifiable. 

The full log-likelihood is
\begin{equation}
L^F\!\left(\{\beta_0^{(k)}\}_{k=1}^K,\{\beta_1^{(k)}\}_{k=1}^K,a\right) = \sum_{r=0}^{1}\sum_{s=1}^{n_r}\log\,l(r,s),
\label{Methodology_FLEX-CP-DT log-Lik}
\end{equation}
and the MLEs $\hat{\beta}_0^{(k)}$, $\hat{\beta}_1^{(k)}$, $\hat{a}$ are obtained via standard numerical optimization.

Note that when $K=1$ (a single interval), the model reduces to the standard negative binomial in~\eqref{Methodology_NB model}.

\subsection{Formula-based conditional power calculation}

Let $\hat{\beta}^*_{1I}$ and $\hat{\beta}^*_{1F}$ denote the estimates of the overall log rate ratio at the interim and final analyses, respectively, both obtained from the same piecewise negative binomial model in \eqref{Methodology_likelihood_factorization}. By standard large-sample theory, their joint distribution is approximately bivariate normal:
\begin{equation}
    \begin{pmatrix}
        \hat{\beta}^*_{1I}\\
        \hat{\beta}^*_{1F}
    \end{pmatrix} \sim N\left[\begin{pmatrix}
            {\beta}^*_{1}\\
        {\beta}^*_{1}
        \end{pmatrix},\begin{pmatrix}
            \sigma^2_{{\beta}^*_{1I}} & \rho_{\beta^*_1}\sigma_{{\beta}^*_{1I}}\sigma_{{\beta}^*_{1F}} \\
            \rho_{\beta^*_1}\sigma_{{\beta}^*_{1I}}\sigma_{{\beta}^*_{1F}} & \sigma^2_{{\beta}^*_{1F}}
        \end{pmatrix}\right],
    \label{methodology bivariate normal dist between Beta_{1I} and Beta_{1F}}
\end{equation}
where the correlation $\rho_{\beta^*_1} = \sigma_{{\beta}^*_{1F}}/\sigma_{{\beta}^*_{1I}}$ follows from the approximate independent-increment assumption \citep{Quan2021}. Given the observed interim estimate $\hat{\beta}^*_{1I}=b^*_{1I}$ and assuming $\beta^*_1 = b^*_{1I}$ (i.e., the unobserved data will follow the interim trend), the conditional power is
\begin{equation}
    \begin{split}
        \text{CP} &= \Pr\left(\hat{\beta}^*_{1F}<-z_{\alpha/2}\,\sigma_{{\beta}^*_{1F}}\;\middle|\;\hat{\beta}^*_{1I}=b^*_{1I}\right) \\
        &= \Phi\left( \frac{-z_{\alpha/2}\,\sigma_{{\beta}^*_{1F}}-b^*_{1I}}{\sigma_{{\beta}^*_{1F}}\sqrt{1-\rho^2_{\beta^*_1}}}\right),
    \end{split}
    \label{methodolody CP formula}
\end{equation}
where $\Phi(\cdot)$ denotes the standard normal cumulative distribution function, $\alpha$ is the two-sided type~I error rate, and $z_{\alpha/2}$ is the upper $(\alpha/2)$-quantile of the standard normal distribution. 

Computing ~\eqref{methodolody CP formula} requires the variances $\sigma^2_{\beta^*_{1I}}$ and $\sigma^2_{\beta^*_{1F}}$. The variance at the final analysis can be computed using the closed form as shown in \citep{Zhu2013}:
\begin{equation}
    \sigma^2_{\beta^*_{1F}} = \frac{1+\theta\, e^{\beta^*_1}}{n_0\,\theta\,\bar{t}\, e^{\beta^*_0+\beta^*_1}} + \frac{(1+\theta)\,a}{n_0\,\theta},
 \label{methodology_var beta_1F}
\end{equation}
where $\theta = n_1/n_0$ denotes the allocation ratio and $\bar{t}$  is the expected average exposure time across all subjects at the final analysis. \textcolor{black}{Notably, using the standard negative binomial variance here is justified by the behavior of the piecewise model at the final
 analysis. When nearly all subjects have completed the planned duration, $t_{rs}=T$ and $K^\prime(r,s)=K$, so the
 total expected count $M_{rs}$ reduces to the arm-level cumulative mean $\mu_r(T)=\Lambda_r T$. The factorization
 in~\eqref{Methodology_likelihood_factorization} then separates into a negative binomial block in the total count
 $y_{rs}$, which carries all information on the overall rate ratio, and a multinomial block governing only the
 temporal allocation of events. Because $y_{rs}$ is sufficient for the overall rate under both the piecewise and the
 standard negative binomial models, the two yield identical inference on $\beta^*_{1F}$, and $\sigma^2_{\beta^*_{1F}}$ coincides with the closed form in~\eqref{methodology_var beta_1F}.}

Under the FLEX-CP-DT framework, the overall log rate ratio at the interim analysis in~\eqref{Methodolog_NB overall efficacy estimation} is expressed as:
\begin{equation}
\hat{\beta}^*_{1I} = \log\left(\frac{\frac{1}{T}\sum_{k=1}^K(T_k - T_{k-1})\cdot\exp\left\{\hat{\beta}_{0I}^{(k)}+\hat{\beta}_{1I}^{(k)}\right\}}{\frac{1}{T}\sum_{k=1}^K(T_k - T_{k-1})\cdot\exp\left\{\hat{\beta}_{0I}^{(k)}\right\}}\right),
\label{Methodolog_NB overall efficacy estimation with FLEX CP DT}
\end{equation}
where $\hat{\beta}_{0I}^{(k)}$ and $\hat{\beta}_{1I}^{(k)}$ denote the interim MLE estimates for the $k$th interval $(T_{k-1},T_k]$. No closed-form expression exists for $\sigma^2_{\beta^*_{1I}}$; instead, we approximate it via the Delta method applied to the observed Hessian matrix $\bm{H}_{\bm{\beta}}^{2K\times 2K}$ from the MLE of~\eqref{Methodology_FLEX-CP-DT log-Lik}.  
It is straightforward to show that 
\begin{equation}
\begin{split}
    \sigma^2_{\beta^*_{1I}} &\approx  \nabla g(\bm{\beta})^\prime\cdot \text{Cov}{(\hat{\bm{\beta}})}\cdot \nabla g(\bm{\beta})\\
    & =\nabla g(\bm{\beta})^\prime\cdot \left(\bm{H}_{\bm{\beta}}^{2K\times 2K}\right)^{-1}\cdot \nabla g(\bm{\beta}),
\end{split}
\label{methodology_var beta_1I}
\end{equation}
where $\nabla g(\bm{\beta})$ is the gradient of $g(\beta) = \hat{\beta}^*_{1I}$ evaluated at  $\hat{\bm{\beta}} = (\hat{\beta}_{0I}^{(1)},\ldots,\hat{\beta}_{0I}^{(K)},\hat{\beta}_{1I}^{(1)},\ldots,\hat{\beta}_{1I}^{(K)})^\prime$.

The proposed FLEX-CP-DT framework is advantageous in that it captures temporal trends in event rates and treatment effects at the interim analysis, while retaining a closed-form conditional power formula~\eqref{methodolody CP formula} through~\eqref{methodology_var beta_1I}, which is computationally efficient. Moreover, in practice, the change points $T_1,\ldots,T_{K-1}$ and the number of intervals $K$ are unknown. The biostatisticians can collaborate with clinicians to pre-specify a set of candidate change-point models that are clinically interpretable for the disease area or based on historical data. The Bayesian Information Criterion (BIC) can then be employed to select the best-fitting model from the candidate models for the CP calculation at the interim analysis.

%% file: Sections/simulation.tex
\section{Simulation}
\label{Simulation}

We conduct simulation studies to assess the performance of the proposed FLEX-CP-DT framework for interim futility decision-making under time-varying treatment effects.

\subsection{Design and data generation}

Consider a pseudo late-phase clinical trial with a recurrent event endpoint as the primary outcome. The main design assumptions are summarized in Table~\ref{simulation_main trial design assumptions}: 1{,}200 patients are uniformly enrolled over an 18-month period and randomized 1:1 between treatment and control arms, each followed for a one-year treatment period ($T=1$). The control arm annualized event rate is $e^{\beta_0^*}=1.5$, the dispersion parameter is $a=1.5$, and an exponential dropout model calibrated to an annual rate of 10\% yields an average exposure of 0.95 years. Under these assumptions, the design achieves approximately 90\% power \citep{Zhu2013} to detect a 25\% reduction in the event rate ($e^{\beta_1^*}=0.75$) at the final analysis. The setup is informed by sample size ranges from historical late-phase trials \citep{castro2018dupilumab,menzies2021tezepelumab,bhatt2023dupilumab,bhatt2024dupilumab}.

\begin{table}[!htb]
\centering
\caption{Trial design assumptions.}
\begin{tabular}{ll}
\toprule
Parameter & Value \\
\midrule
Annualized event rate for control group & $e^{\beta^*_0} = 1.5$\\
Annualized event rate for treatment group & $e^{\beta^*_0 + \beta^*_1} = 1.125$ \\
Fixed treatment period & $T=1$ year\\
Dispersion parameter & $a=1.5$ \\
Two-sided type I error & $\alpha =0.05$ \\
Dropout rate per year & $r=0.1$ \\
Uniform enrollment period & 18 months\\
\bottomrule
\end{tabular}
\label{simulation_main trial design assumptions}
\end{table}

To examine the impact of time-varying event rates on interim futility decisions and to highlight the advantages of FLEX-CP-DT, we consider four scenarios with an overall 25\% risk reduction at the final analysis ($e^{\beta_1^*}=0.75$). These scenarios fall into two categories defined by the temporal pattern of the treatment effect: (i)~\emph{constant-effect} (Scenario~1), in which the event rate ratio is time-invariant over the treatment period; and (ii)~\emph{delayed-effect} (Scenarios~2--4), in which the treatment exerts no benefit early but a pronounced benefit later, with the cumulative effect over $[0,T]$ averaging to a 25\% reduction. Table~\ref{simulation_scenario_summary} provides the per-scenario specifications.

\begin{table}[!htb]
\centering
\caption{Overview of the four simulation scenarios. The control event rate is $e^{\beta_0^*}=1.5$ throughout.}
\begin{tabular}{clll}
\toprule
Scenario & Category & Change points & $e^{\beta_1(t)}$ profile \\
\midrule
1 & Constant effect & none & $0.75$ \\
2 & Delayed onset (three segments) & $T/3,\,2T/3$ & $1.00 \to 0.80 \to 0.45$ \\
3 & Delayed onset (at one-third) & $T/3$ & $1.00 \to 0.625$ \\
4 & Delayed onset (at one-quarter) & $T/4$ & $1.00 \to 0.667$ \\
\bottomrule
\end{tabular}
\label{simulation_scenario_summary}
\end{table}

For each scenario, event data were generated from the conditional Poisson process~\eqref{Methodology_conditional_Poisson} with subject-specific gamma frailties $\nu_{rs}\sim\text{Gamma}(a^{-1},a^{-1})$. Specifically, individual event occurrence times were drawn from a piecewise-constant intensity $\nu_{rs}\cdot\exp\{\beta_0^{(k)}+r\cdot\beta_1^{(k)}\}$ on each interval $(T_{k-1},T_k]$, where $\beta_0^{(k)}$ and $\beta_1^{(k)}$ encode the scenario-specific temporal pattern. Integrating out the frailty yields marginal negative binomial counts with dispersion $a=1.5$. Dropout times were generated independently from an exponential distribution. Interim analyses were performed when 30\%, 40\%, and 50\% of patients had completed at least 24 weeks of follow-up at the corresponding data cut-off. Each scenario was replicated 10{,}000 times.

\subsection{Methods compared}

At each interim analysis, FLEX-CP-DT was fitted using a pre-specified set of five candidate piecewise working models on $[0,T]$:
\begin{enumerate}
    \item change points at $T/3$ and $2T/3$ (three equal segments);
    \item change point at $T/2$ (two equal segments);
    \item change points at $T/4$, $T/2$, and $3T/4$ (four equal segments);
    \item change point at $T/3$ (two segments);
    \item change point at $T/4$ (two segments).
\end{enumerate}
The working model for conditional power calculation was selected by BIC. FLEX-CP-DT was compared against four variants of Hui's method \citep{Quan2021}, which apply the standard negative binomial model~\eqref{Methodology_NB model} to progressively restricted subsets of the interim cohort. Specifically, Hui1--Hui4 require patients to have observed treatment durations of at least 0, 3, 6, and 9 months, respectively. For each method, we computed the interim point estimate $\hat{\beta}^*_{1I}$, the corresponding variance $\sigma^2_{\beta^*_{1I}}$, the projected final-analysis variance $\sigma^2_{\beta^*_{1F}}$ under the design assumptions, and the resulting conditional power via~\eqref{methodolody CP formula} at a two-sided significance level of $\alpha=0.05$.

Two futility criteria were evaluated. The first claims futility when the conditional power falls below 20\%. The second claims futility when the observed interim event rate ratio $\exp(\hat{\beta}^*_{1I})$ exceeds 0.85. In addition to futility-decision performance, we report the root mean square error (RMSE) of $\hat{\beta}^*_{1I}$ relative to the target estimand $\beta^*_1$ to assess the precision of each method's interim estimator.

\subsection{Results}

We evaluate operating characteristics under two perspectives: the probability of incorrectly claiming futility at the interim analysis (Tables~\ref{simulation_conditional power summary} and~\ref{simulation_observed IA futility summary}) and the precision of the underlying interim estimator (Table~\ref{simulation_RMSE summary}). Because all four scenarios have a true 25\% risk reduction at the final analysis, a method performs well when it claims futility infrequently---particularly under the delayed-effect profiles where interim data may misleadingly suggest no benefit.

Under the delayed-onset profiles (Scenarios~2--4), FLEX-CP-DT consistently yields a lower probability of incorrectly claiming futility than Hui1--Hui4 across both criteria and all three interim analysis timings. In Scenario~1, where the constant-rate assumption holds and the standard model is correctly specified,  FLEX-CP-DT performs slightly worse than Hui1 (though the difference is negligible), comparably to Hui2, and slightly better than Hui3 and Hui4.

Three general patterns hold across the four scenarios. First, the probability of incorrectly claiming futility decreases monotonically as the proportion of 24-week completers increases from 30\% to 50\% across all methods, reflecting the increasing precision of the interim estimations with more mature data. Second, at a fixed interim analysis timing, the four Hui variants do not exhibit a monotone pattern in RMSE across the delayed-effect scenarios except in the constant-effect Scenario~1. This reflects two competing factors: tighter minimum-follow-up thresholds reduce the bias contributed by patients with limited exposure but also shrink the effective sample size, inflating variance. The trade-off produces a non-monotone RMSE pattern. For example, in Scenario~2 when 30\% have completed 24 weeks, RMSEs are 0.240, 0.233, 0.230, 0.300 for Hui1--Hui4, respectively. FLEX-CP-DT addresses this trade-off by modeling the temporal pattern directly rather than restricting the interim cohort, thereby making full use of the data available at the interim analysis. Third, across replicates, the BIC-selected working model was most often candidate~(2), the single change point at $T/2$. This is consistent with BIC's preference for parsimony: a single mid-interval change point is often adequate to capture the dominant temporal feature in Scenarios~2--4.

Overall, FLEX-CP-DT behaves comparably to the standard approach when the constant-rate assumption holds and substantially reduces erroneous futility claims when the treatment effect exhibits delayed onset.

\begin{table}[!htb]
\centering
\caption{Probability of incorrectly claiming interim futility based on conditional power $< 20\%$.}
\begin{tabular}{ccccccc}
\toprule
&&&&Method\\ \cline{3-7} \rule{0pt}{10pt}
 Scenario & \% completed 24 weeks & Hui1\textsuperscript{1} & Hui2 & Hui3 & Hui4 & FLEX-CP-DT\\
\midrule
\multirow{3}{*}{1} & 30\% & 0.111\textsuperscript{2} & 0.120 & 0.157 & 0.269 & 0.138 \\
& 40\% & 0.085 & 0.095 & 0.122 & 0.183 & 0.100 \\
& 50\% & 0.073 & 0.081 & 0.100 & 0.136 & 0.082 \\ \midrule
\multirow{3}{*}{2} & 30\% & 0.553 & 0.506 & 0.406 & 0.344 & 0.298 \\
& 40\% & 0.480 & 0.428 & 0.327 & 0.256 & 0.209 \\
& 50\% & 0.411 & 0.354 & 0.252 & 0.182 & 0.153 \\ \midrule
\multirow{3}{*}{3} & 30\% & 0.435 & 0.371 & 0.276 & 0.302 & 0.189 \\
& 40\% & 0.352 & 0.294 & 0.207 & 0.205 & 0.144 \\
& 50\% & 0.300 & 0.243 & 0.163 & 0.157 & 0.115 \\ \midrule
\multirow{3}{*}{4} & 30\% & 0.334 & 0.273 & 0.229 & 0.281 & 0.194 \\
& 40\% & 0.283 & 0.227 & 0.175 & 0.196 & 0.151 \\
& 50\% & 0.232 & 0.187 & 0.143 & 0.158 & 0.120 \\
\bottomrule
\multicolumn{7}{p{13cm}}{\small \textsuperscript{1}Hui's method is applied to four subsets of the interim data with minimum observed durations of 0, 3, 6, and 9 months, respectively. \newline
\textsuperscript{2}Based on 10{,}000 replications; each entry is the proportion of replicates with conditional power below 20\%.}
\end{tabular}
\label{simulation_conditional power summary}
\end{table}

\begin{table}[!htb]
\centering
\caption{Probability of incorrectly claiming interim futility based on observed rate ratio $\exp(\hat{\beta}^*_{1I})>0.85$.}
\begin{tabular}{ccccccc}
\toprule
&&&&Method\\ \cline{3-7} \rule{0pt}{10pt}
 Scenario & \% completed 24 weeks & Hui1\textsuperscript{1} & Hui2 & Hui3 & Hui4 & FLEX-CP-DT\\
\midrule
\multirow{3}{*}{1} & 30\% & 0.188\textsuperscript{2} & 0.202 & 0.241 & 0.342 & 0.220 \\
                            & 40\% & 0.159 & 0.174 & 0.202 & 0.272 & 0.182 \\
                            & 50\% & 0.137 & 0.143 & 0.174 & 0.220 & 0.151 \\ \midrule
\multirow{3}{*}{2} & 30\% & 0.690 & 0.639 & 0.530 & 0.434 & 0.417 \\
                            & 40\% & 0.621 & 0.570 & 0.452 & 0.352 & 0.318 \\
                            & 50\% & 0.548 & 0.489 & 0.375 & 0.278 & 0.245 \\ \midrule
\multirow{3}{*}{3} & 30\% & 0.574 & 0.511 & 0.388 & 0.383 & 0.294 \\
                            & 40\% & 0.488 & 0.427 & 0.313 & 0.298 & 0.240 \\
                            & 50\% & 0.428 & 0.366 & 0.263 & 0.246 & 0.200 \\ \midrule
\multirow{3}{*}{4} & 30\% & 0.469 & 0.401 & 0.336 & 0.359 & 0.290 \\
                            & 40\% & 0.409 & 0.343 & 0.276 & 0.283 & 0.246 \\
                            & 50\% & 0.353 & 0.290 & 0.233 & 0.241 & 0.205 \\
\bottomrule
\multicolumn{7}{p{13cm}}{\small \textsuperscript{1}Hui's method is applied to four subsets of the interim data with minimum observed durations of 0, 3, 6, and 9 months, respectively. \newline
\textsuperscript{2}Based on 10{,}000 replications; each entry is the proportion of replicates with $\exp(\hat{\beta}^*_{1I})>0.85$.}
\end{tabular}
\label{simulation_observed IA futility summary}
\end{table}

\begin{table}[!htb]
\centering
\caption{RMSE of the interim estimator $\hat{\beta}^*_{1I}$.}
\begin{tabular}{ccccccc}
\toprule
&&&&Method\\ \cline{3-7} \rule{0pt}{10pt}
 Scenario & \% completed 24 weeks & Hui1\textsuperscript{1} & Hui2 & Hui3 & Hui4 & FLEX-CP-DT\\
 \midrule
\multirow{3}{*}{1} & 30\% & 0.141\textsuperscript{2} & 0.151 & 0.184 & 0.297 & 0.166 \\
                            & 40\% & 0.125 & 0.133 & 0.153 & 0.205 & 0.138 \\
                            & 50\% & 0.114 & 0.120 & 0.134 & 0.163 & 0.121 \\
\midrule
\multirow{3}{*}{2} & 30\% & 0.240 & 0.233 & 0.230 & 0.300 & 0.187 \\
                            & 40\% & 0.205 & 0.197 & 0.186 & 0.209 & 0.148 \\
                            & 50\% & 0.180 & 0.172 & 0.158 & 0.169 & 0.128 \\
\midrule
\multirow{3}{*}{3} & 30\% & 0.205 & 0.198 & 0.198 & 0.298 & 0.168 \\
                            & 40\% & 0.174 & 0.166 & 0.162 & 0.204 & 0.139 \\
                            & 50\% & 0.154 & 0.146 & 0.141 & 0.166 & 0.122 \\
\midrule
\multirow{3}{*}{4} & 30\% & 0.181 & 0.173 & 0.188 & 0.292 & 0.167 \\
                            & 40\% & 0.159 & 0.152 & 0.158 & 0.204 & 0.140 \\
                            & 50\% & 0.140 & 0.133 & 0.138 & 0.168 & 0.123 \\
\bottomrule
\multicolumn{7}{p{13cm}}{\small \textsuperscript{1}Hui's method is applied to four subsets of the interim data with minimum observed durations of 0, 3, 6, and 9 months, respectively. \newline
\textsuperscript{2}Based on 10{,}000 replications; $\text{RMSE}=\sqrt{(1/10{,}000)\sum(\hat{\beta}^*_{1I}-\beta_1^*)^2}$.}
\end{tabular}
\label{simulation_RMSE summary}
\end{table}

%% file: Sections/application.tex
\section{Application}
\label{Application}

\subsection{Background}

Bronchiectasis is a chronic lung condition characterized by abnormal widening of the airways, leading to excessive mucus buildup and recurrent infections. A pulmonary exacerbation (PE) is defined as the presence of at least three respiratory symptoms for at least 24 hours that prompts a physician to prescribe systemic antibiotics; reduction in the annualized PE rate is the primary efficacy endpoint.

Our application is motivated by the ASPEN study (NCT04594369), a phase 3 trial of brensocatib (a DPP-1 inhibitor) in patients with bronchiectasis \citep{chalmers2025phase}. In this double-blind trial, patients were randomly assigned to receive brensocatib (10 mg or 25 mg once daily) or placebo over a 52-week treatment period. The primary efficacy readout is reproduced in Table~\ref{Application_ASPEN readout}.

\begin{table}[!htb]
\centering
\caption{Primary Endpoints (Intention-to-Treat Population)}
\begin{tabular}{lccc}
\toprule
 Primary Endpoint                                 & \makecell{ Brensocatib, 10mg\\ (N=583)}         &	\makecell{Brensocatib, 25mg \\(N=575)}&	\makecell{Placebo\\(N=563)}  \\ \midrule
\makecell[l]{Annualized rate of \\pulmonary exacerbations\\ — no. of events/yr  (95\% CI)\\} &	\makecell{1.02 \\(0.91 to 1.13)}	&\makecell{1.04\\(0.93 to 1.16)}&\makecell{1.29\\(1.16 to 1.43)}\\ [3ex]
Rate ratio (95\%) &	\makecell{0.79\\(0.68 to 0.92)}	&\makecell{0.81\\(0.69 to 0.94)}	&\makecell[c]{\textemdash{}}\\[3ex]
Adjusted P value&	0.004&	0.005&	\makecell[c]{\textemdash{}}\\

\bottomrule
\end{tabular}

\label{Application_ASPEN readout}
\end{table}

\begin{figure}[!htb]
    \centering
    \includegraphics[width=.8\linewidth]{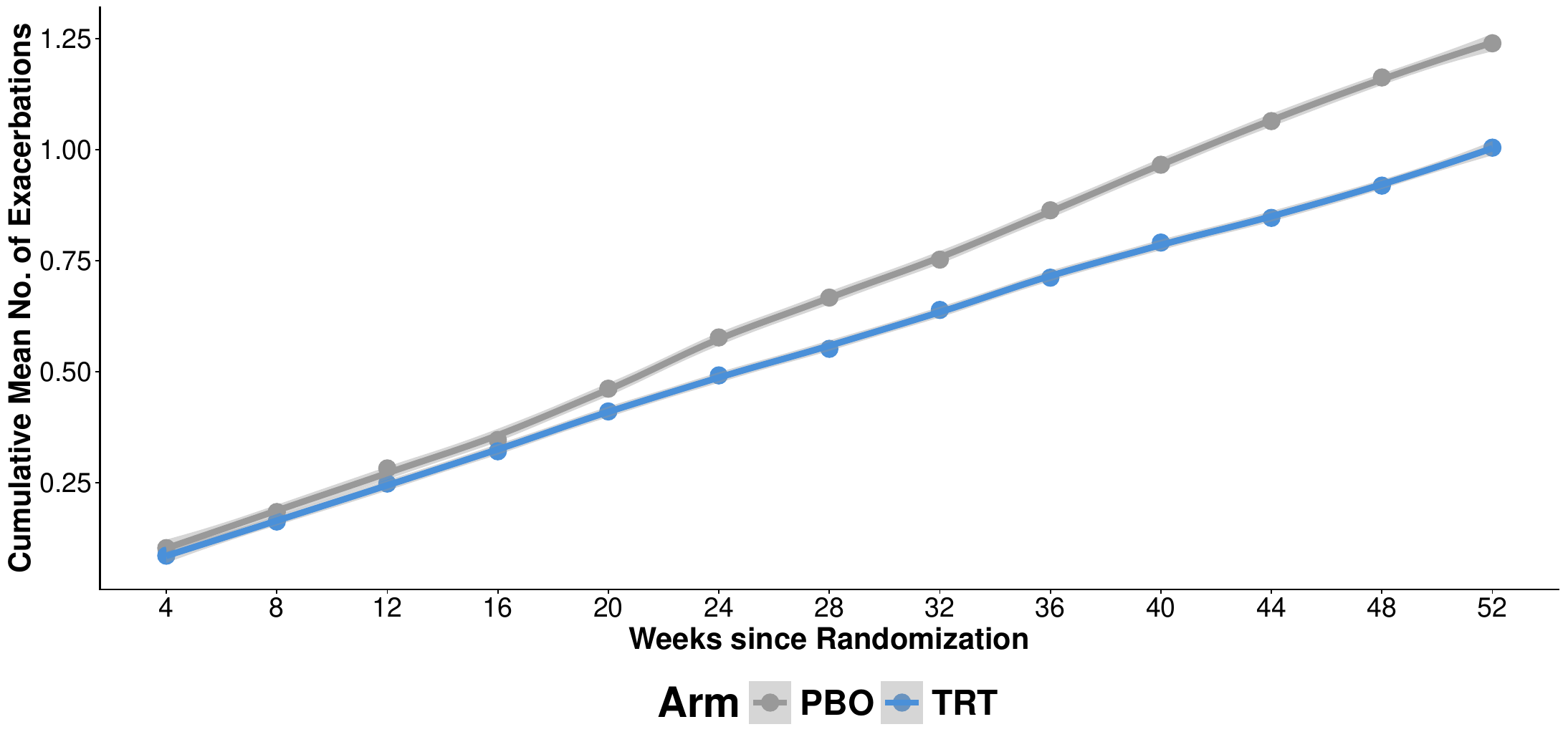}
    \caption{Simplified Mean Cumulative Functions (MCFs) of pulmonary exacerbations (PEs) between placebo (PBO) and treatment (TRT) arms over 52 weeks}
    \label{Application_ASPEN MCF}
\end{figure}

Although the DPP-1 inhibitor demonstrated statistical significance ($P \le 0.01$) on an approximately 20\% PE rate reduction at the final analysis, the Mean Cumulative Functions (MCFs) of PEs suggests no immediate drug effect: the curves overlap through Week 20 and diverge in a sustained manner thereafter, implying a delayed treatment effect. We digitized the published MCF curves using WebPlotDigitizer and present a simplified version of the delayed-effect pattern in Figure~\ref{Application_ASPEN MCF}; all subsequent calculations are based on this simplified pattern.

\begin{figure}[!htb]
    \centering
    \includegraphics[width=\linewidth]{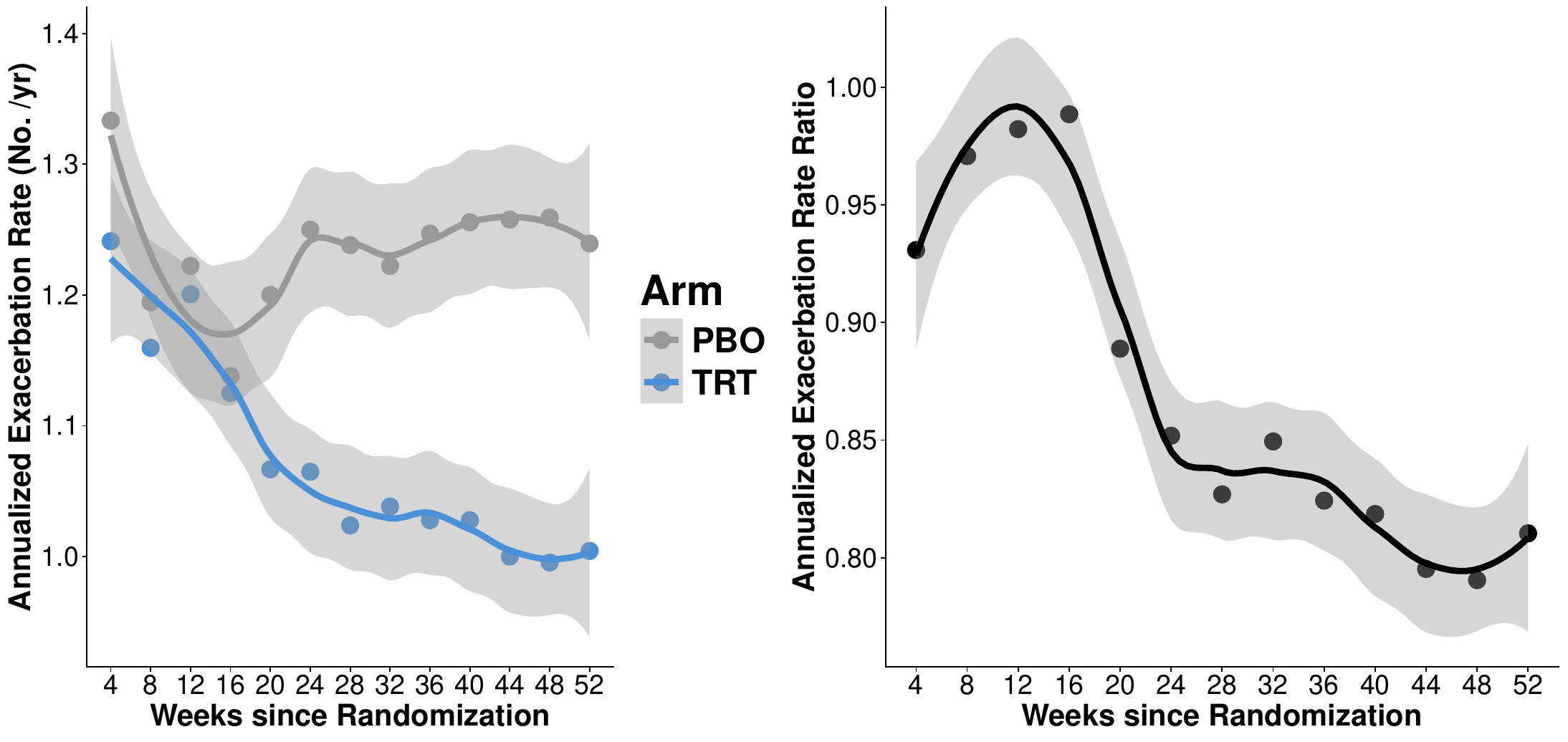}
    \caption{Annualized PE rate over 52 weeks (left) and PE rate ratio over 52 weeks (right)}
    \label{Application_ASPEN annualized PE rate and RR}
\end{figure}

From the MCF curves, the annualized PE rate and rate ratio over time are derived in Figure~\ref{Application_ASPEN annualized PE rate and RR}. Before Week 20, the rates in both arms hover around 1.1--1.3 PEs/year with the rate ratio near 1; a 20\% rate reduction does not emerge until patients complete Week 52. This pattern motivates the question: how should interim futility decisions be made when the investigational drug exhibits a delayed treatment effect?

\subsection{Calibrated simulation design}

To evaluate FLEX-CP-DT against Hui's method in a clinically realistic setting, we conducted a simulation study calibrated to the ASPEN readout. Each replicate reproduces the design of the reference study: 575 patients randomized to treatment and 563 to placebo, a 28-month enrollment period with monthly enrollment fractions specified in Table~\ref{Application_Monthl Enrollment Rates}, a 52-week treatment period, and an annual dropout rate of 10\%.

\begin{table}[h]
    \centering
    \begin{tabular}{c|c}
    \toprule
        Enrollment Months & Monthly enrollment rates\\
        \midrule
        1-4 & 1\%, 2\%, 3\%,4\% (respectively)\\
        5-17 & 5\% \\
        18-20 & 4\% \\
        21 & 3\% \\
        22-24 & 2\%\\
        25-28 & 1\%\\
    \bottomrule
    \end{tabular}
    \caption{Monthly enrollment setting}
    \label{Application_Monthl Enrollment Rates}
\end{table}

Based on the delayed-effect pattern in Figures~\ref{Application_ASPEN MCF} and~\ref{Application_ASPEN annualized PE rate and RR}, we assume piecewise-constant annualized exacerbation rates with a single change point at approximately Week 18 ($t=0.346$ on the unit interval). The piecewise rates are summarized in Table~\ref{Application_piecewise exacerbation rates}; the dispersion parameter is set to 1.2.

\begin{table}[h]
    \centering
    \begin{tabular}{ccc}
    \toprule
         & $[0,\,0.346]$ & $(0.346,\,1]$\\
        \midrule
        Placebo   & 1.18  & 1.25  \\
        Treatment & 1.10  & 0.948 \\
    \bottomrule
    \end{tabular}
    \caption{Piecewise annualized exacerbation rates in the placebo and treatment arms over the 52-week treatment period, scaled to the unit interval $[0,1]$.}
    \label{Application_piecewise exacerbation rates}
\end{table}

To ensure each replicate is representative of the reference trial, we applied a calibration check: after generating a trial, we fit a standard negative binomial model to the complete simulated data, and retained the replicate only if the estimated treatment rate, control rate, rate ratio, and confidence interval width all fell within 5\% of the published ASPEN values. For each qualified replicate, interim analyses were conducted when 30\%, 40\%, and 50\% of patients had completed at least 24, 36, or 52 weeks of follow-up. FLEX-CP-DT was fitted with the same five candidate models as in Section~\ref{Simulation}, and compared against Hui1--Hui4. Two futility criteria were applied: conditional power below 20\%, and observed interim rate ratio $\exp(\hat{\beta}^*_{1I})$ exceeding 0.85.

\subsection{Results}

Tables~\ref{application_conditional power summary} and~\ref{application_observed IA futility summary} report the probability of claiming interim futility under the CP-based approach and the effect size based approach, respectively.

\begin{table}[!htb]
\centering
\caption{Comparison of conditional power-based futility decisions between FLEX-CP-DT framework and Hui's method}
\begin{tabular}{ccccccc}
\toprule
&&&&Method\\ \cline{3-7} \rule{0pt}{10pt}
 Completion Week & \% completed & Hui1\textsuperscript{1} & Hui2 & Hui3 & Hui4 & FLEX-CP-DT\\
\midrule
\multirow{3}{*}{24} & 30\% & 0.338\textsuperscript{2} & 0.319 & 0.287 & 0.311 & 0.24 \\
  & 40\% & 0.27 & 0.25 & 0.222 & 0.24 & 0.176 \\
  & 50\% & 0.21 & 0.194 & 0.17 & 0.182 & 0.132 \\ \midrule
  \multirow{3}{*}{36} & 30\% & 0.242 & 0.226 & 0.202 & 0.216 & 0.157 \\
  & 40\% & 0.191 & 0.178 & 0.155 & 0.167 & 0.118 \\
  & 50\% & 0.148 & 0.137 & 0.117 & 0.126 & 0.089 \\ \midrule
  \multirow{3}{*}{52} & 30\% & 0.152 & 0.145 & 0.124 & 0.131 & 0.091 \\
  & 40\% & 0.115 & 0.107 & 0.092 & 0.101 & 0.068 \\
  & 50\% & 0.077 & 0.075 & 0.062 & 0.075 & 0.047 \\
\bottomrule
\multicolumn{7}{p{14cm}}{\small \textsuperscript{1}Hui's method is applied to four subsets of the interim data with minimum observed durations of 0, 3, 6, and 9 months, respectively. \newline
\textsuperscript{2}Based on 10{,}000 replications; each entry is the proportion of replicates with conditional power below 20\%.}
\end{tabular}
\label{application_conditional power summary}
\end{table}

\begin{table}[!htb]
\centering
\caption{Comparison of observed interim efficacy-based futility decisions between FLEX-CP-DT framework and Hui's method}
\begin{tabular}{ccccccc}
\toprule
&&&&Method\\ \cline{3-7} \rule{0pt}{10pt}
 Completion Week & \% completed & Hui1\textsuperscript{1} & Hui2 & Hui3 & Hui4 & FLEX-CP-DT\\
\midrule
\multirow{3}{*}{24} & 30\% & 0.517\textsuperscript{2} & 0.489 & 0.421 & 0.413 & 0.367 \\
  & 40\% & 0.449 & 0.427 & 0.378 & 0.37 & 0.318 \\
  & 50\% & 0.394 & 0.369 & 0.324 & 0.32 & 0.272 \\ \midrule
  \multirow{3}{*}{36} & 30\% & 0.43 & 0.406 & 0.356 & 0.347 & 0.298 \\
  & 40\% & 0.377 & 0.358 & 0.302 & 0.304 & 0.258 \\
  & 50\% & 0.318 & 0.3 & 0.265 & 0.267 & 0.214 \\ \midrule
  \multirow{3}{*}{52} & 30\% & 0.324 & 0.311 & 0.273 & 0.276 & 0.219 \\
  & 40\% & 0.266 & 0.256 & 0.229 & 0.236 & 0.185 \\
  & 50\% & 0.204 & 0.196 & 0.179 & 0.196 & 0.14 \\
\bottomrule
\multicolumn{7}{p{14cm}}{\small \textsuperscript{1}Hui's method is applied to four subsets of the interim data with minimum observed durations of 0, 3, 6, and 9 months, respectively. \newline
\textsuperscript{2}Based on 10{,}000 replications; each entry is the proportion of replicates with observed interim rate ratio $\exp(\hat{\beta}^*_{1I})$ exceeding 0.85.}
\end{tabular}
\label{application_observed IA futility summary}
\end{table}

Across all configurations, FLEX-CP-DT yields a lower probability of incorrectly claiming futility than all four Hui variants. At the earliest cut-off (30\% of patients completed week 24), the CP-based approach futility probability is 0.240 for FLEX-CP-DT versus 0.287-0.338 for Hui1 - Hui4, and the effect size based approach gives futility probability of 0.367 versus 0.413 - 0.517. The gap narrows as the interim analysis was performed with more mature data. The four Hui variants do not exhibit a strict monotone ordering across the minimum-follow-up threshold, reflecting the same bias-variance trade-off discussed in Section~\ref{Simulation}.

Overall, the calibrated case study confirms the simulation findings: when the treatment effect is delayed, FLEX-CP-DT reduces the probability of falsely terminating an efficacious drug at the interim analysis.

%% file: Sections/discussion.tex
\section{Discussion}
\label{Discussion}

In this study, we propose the FLEX-CP-DT framework for interim futility decision-making in clinical trials with recurrent event endpoints. The framework employs a piecewise-constant rate model with shared gamma frailty. Its marginal likelihood combines a negative binomial component for the total event counts with a multinomial component for temporal allocation of events across intervals, enabling identification of time-varying event rates and treatment effects at the interim analysis. The resulting conditional power formula retains a closed form that is computationally efficient. By modeling the time-varying structure directly, FLEX-CP-DT addresses a key limitation of the standard negative binomial model with an offset for patient follow-up time: at the interim analysis, when a substantial fraction of patients have incomplete follow-up, the constant-rate assumption can bias the estimated treatment effect and lead to incorrect futility decisions.

The simulation study demonstrates that under delayed-onset treatment effects, FLEX-CP-DT substantially reduces the probability of erroneously claiming futility compared to Hui's method, which applies the standard negative binomial model to progressively restricted subsets of the interim cohort. Under the constant-effect scenario, FLEX-CP-DT performs comparably to the standard approach. The real case study based on a bronchiectasis trial with a DPP-1 inhibitor further confirms these findings in a clinically realistic setting: FLEX-CP-DT reduces the probability of falsely terminating an efficacious drug whose benefit emerges after an initial period of no observable treatment effect.

A further practical advantage of the piecewise-constant model is that a set of candidate working models can be pre-specified in the study protocol based on collaboration between statisticians and clinicians. Change-point locations can be motivated by historical data or clinical knowledge of the drug's mechanism of action or expected time to treatment onset. By contrast, pre-specifying a continuous functional form---such as a spline-based intensity model---before observing any data is more challenging, as the choice of knot locations and smoothing parameters is difficult to justify a priori without data. \textcolor{black}{One potential concern is that zero events may be observed in certain intervals, which would render the corresponding $\beta_1^{(k)}$ non-estimable. Such a scenario would most likely occur under a granular candidate model, such as a specification with weekly change points. In practice, however, a parsimonious model is typically sufficient to capture the dominant temporal trend, and hence the likelihood of zero events in certain intervals should be ignorable.}

It is noteworthy that the proposed framework is designed specifically for the interim analysis stage where heterogeneous follow-up durations introduce bias under the constant-rate assumption. At the final analysis, when nearly all patients have completed the planned treatment duration, the standard negative binomial model with an offset to adjust for follow-up time remains appropriate, as the treatment effect estimated by the standard model is interpretable as the ratio of cumulative mean event counts over the planned treatment period regardless of the temporal pattern.

Several extensions of the FLEX-CP-DT framework warrant investigation in future work. First, the negative binomial model may not be appropriate when its Poisson-gamma distributional assumptions are violated, as recently demonstrated by \cite{Paulson27052026} under non-gamma between-subject heterogeneity. While the Andersen-Gill model with robust variance remains robust in such settings, it is anticipated to face the same non-constant intensity challenge at interim analysis as the standard negative binomial model. Extending the FLEX-CP-DT approach to an Andersen-Gill-based working model would naturally fill this gap. Second, the current framework does not incorporate baseline covariates; including covariates correlated with the primary endpoint may improve estimation precision and the accuracy of futility decisions. Both extensions require substantial methodological development and will be pursued in separate studies.